\documentclass[fleqn,usenatbib]{mnras}

\usepackage{txfonts}

\usepackage[T1]{fontenc}

\DeclareRobustCommand{\VAN}[3]{#2}
\let\VANthebibliography\thebibliography
\def\thebibliography{\DeclareRobustCommand{\VAN}[3]{##3}\VANthebibliography}

\usepackage{graphicx}	
\usepackage{amssymb}	
\usepackage{multirow}
\usepackage{multicol}
\usepackage{verbatim}
\usepackage{subcaption}
\usepackage{booktabs, caption, makecell}
\usepackage{threeparttable}

\newcommand{\mgr}{\,CXOU~J010043}	
\graphicspath{{Figures/}}

\title[Spectroscopy and timing of CXOU J010043.1$-$721134]{Long-term \textit{XMM-Newton} view of magnetar CXOU J010043.1$-$721134: Comprehensive spectral and temporal results}

\author[Rwitika et al.]{
Rwitika Chatterjee,$^{1}$\thanks{E-mail: rwitika@ursc.gov.in}
Vivek K. Agrawal,$^{1}$
Anuj Nandi,$^{1}$
\\
$^{1}$Space Astronomy Group, ISITE Campus, U. R. Rao Satellite Centre, ISRO, Bengaluru 560037, India\\
}

\date{Accepted XXX. Received YYY; in original form ZZZ}

\pubyear{2020}

\begin{document}
\label{firstpage}
\pagerange{\pageref{firstpage}--\pageref{lastpage}}
\maketitle

\begin{abstract}
We present an in-depth analysis and results of eleven \textit{XMM-Newton} datasets, spanning 2000 to 2016, of the anomalous X-ray Pulsar CXOU~J010043.1$-$721134 which has been classified as a magnetar. We find a spin-period of 8.0275(1)~s as of December 2016 and calculate the period derivative to be $(1.76\pm 0.02) \times 10^{-11}$~s~s$^{-1}$, which translate to a dipolar magnetic field strength of $3.8\times 10^{14}$~G and characteristic age of $\sim 7200$~yr for the magnetar. It has a double-peaked pulse profile, with one broad and one narrow peak, in both soft ($0.3-1.3$~keV) and hard ($1.3-8$~keV) energy bands. The pulse fractions in the two energy bands are found to be consistent with constant values. These results are in agreement with previously published results for this source. Although two-component models produce acceptable fits to its energy spectra, single component models are much simpler and are able to explain the similarity of the pulse profiles in the low and high energy bands. We attempt fitting with four different single-component models and find that the best fit to the spectra is obtained by fitting a thermal Comptonization model with the photon index $(\Gamma)$ between $2.0-2.7$ and the electron temperature $(kT_e)$ between $0.5-0.9$~keV, for a seed blackbody photon distribution of 0.2~keV. Finally, we conclude by discussing our results briefly.
\end{abstract}




\section{Introduction}
\label{sec:intro}

Magnetars are a class of isolated neutron stars which are now known to be powered by the decay of their ultrastrong magnetic fields ($\sim 10^{13}-10^{15}$~G). Historically, two classes of X-ray sources were discovered, Soft Gamma Repeaters (SGRs) and Anomalous X-ray Pulsars (AXPs), which had observationally distinct properties. SGRs were a class of repeatedly bursting X-ray sources, which were seen to have softer burst spectra than the classical Gamma Ray Bursts. AXPs were different from the `conventional' pulsars due to their long pulse periods, steady spin-down, unusually soft X-ray spectra, lack of optical counterparts and higher luminosity than spin-down power. \citet{gavriil2002} first reported SGR-like bursts from the AXP 1E~1048.1$-$5937, followed closely by \citet{kaspi2003} for AXP 1E 2259+586. Moreover, the persistent counterparts of the SGRs were found to show many of the characteristics of AXPs. These findings led to the unification of SGRs and AXPs as a single class of source -- magnetars (see \citealt{mereghetti2008} and \citealt{kaspi2017} for reviews on magnetars). 
\par Magnetars are characterized by high-persistent X-ray luminosity ($\sim 10^{34}-10^{36}$~erg~s$^{-1}$) which is typically two orders of magnitude higher than their spin-down luminosity. In the first known magnetar sources, pulse periods measured at different epochs \citep{baykal1996} clearly demonstrated regular spin-down, attributed to magnetic braking. The possible scenario of formation of such highly magnetized neutron stars was proposed by \citet{duncan1992} (see also \citealt{thompson1995} and \citealt{thompson1996}). As of today, there are 24 confirmed magnetars of which 12 are AXPs and 12 are SGRs, and 6 unconfirmed magnetar candidates \citep{olausen2014}.
\par X-ray pulsations from magnetars are usually in the range $2-12$~s, with large spin-down rates $\dot{P}\sim 10^{-13}-10^{-11}$~s~s$^{-1}$ \citep[see][]{turolla2015} which suggest that these are extremely young objects (spin-down time scale $\sim P/\dot{P}$ in the range 1 to 100~kyr). Many magnetars exhibit a variety of transient behaviours such as short bursts, large outbursts (see \citealt{rea2011}), glitches (sudden spin-up, see \citealt{kaspi2000}), anti-glitches and giant flares in rare cases (see \citealt{mazets1979_2, hurley1999, hurley2005}). The persistent soft X-ray emission from magnetars generally has broad pulse profiles, with one (e.g. 1E~1048.1$-$5937 \citealt{tam2008}, XTE~J1810$-$197 \citealt{bernardini2009}) or two (4U~0142+0162 \citealt{rea2007}) peaks. The X-ray pulse fraction varies greatly from source to source, ranging from 10\% \citep{israel1999} to 70\% \citep{rea2013}, and can be strongly energy-dependent. The pulse profile may also vary with energy and in some cases, have long-term temporal variations \citep{gonzalez2010}.
\par The soft X-ray spectra of magnetars are usually well-fitted by an absorbed blackbody model ($kT\sim 0.3-0.5$~keV) for surface thermal continuum emission which dominates the spectra at lower energies, plus a power-law tail (photon index $\sim 3-4$) for emission from non-thermal processes in the magnetosphere such as inverse-Compton scattering or synchrotron radiation, which becomes significant $\gtrsim$~3~keV (see \citealt{olausen2014} for a compilation of spectral fits to all known magnetars using this model). Often, double blackbodies produce acceptable fits (\citealt{halpern2005}; \citealt{tiengo2008}, hereafter TE08) and some authors have also fitted Comptonized blackbody models to magnetar spectra (see \citealt{tendulkar2015} for 4U~0142+61). \citet{rea2008} presented a resonant cyclotron scattering (RCS) model which describes the resonant scattering of the thermal emission from magnetar surface by hot magnetospheric plasma -- they were able to fit the soft spectra (below 10~keV) of ten magnetars using this model, out of the fifteen known at the time of publishing\footnote{The remaining five were excluded from their analysis due to lower-than-desired signal-to-noise ratio of the observations.}. 
\par Being comparatively young objects, the spatial distribution of magnetars is largely confined near the Galactic plane, with a scale height of only $20-30$~pc \citep{olausen2014}. As a result, the observations of their X-ray spectrum are subject to large interstellar absorption. However, CXOU~J010043.1$-$721134 (hereafter \mgr), an AXP and the only known magnetar in the SMC till date, has the lowest column density ($\sim 10^{20}-10^{21}$~cm$^{-2}$) among all known magnetars, owing to its location away from the Galactic plane. Since the distance to SMC and its absorption column are well-constrained, this source can be investigated with higher accuracy compared to other magnetars.  
\par This source has been serendipitously observed by many X-ray imaging satellites (e.g. \textit{Einstein, ROSAT, ASCA, Chandra}), with observations dating as far back as 1979. However, \citet{lamb2002} first proposed that this source, having an average X-ray luminosity of $1.5\times 10^{34}$~erg~s$^{-1}$ and a soft X-ray spectrum well-fitted by a blackbody model with $kT=(0.41\pm 0.01)$~keV, is an AXP. Although the authors initially found a spin period of 5.44~s for this source using \textit{Chandra ACIS-I} data of May~2001, \citet{lamb2003} and later, \citet{majid2004} corrected it to 8.02~s using newer data for this source from \textit{XMM-Newton} EPIC-PN observations. \citet{majid2004} fitted the source spectrum with power-law as well as a combination of power-law and blackbody components, although they deemed the latter to be a poorer fit than the former.
\par \citet{mcgarry2005} (hereafter MG05) analyzed newer \textit{Chandra} data of \mgr$\,$ (taken in 2004) and found a spin period of 8.02~s, consistent with previous studies. The authors combined the current observations with the 2001 observations to calculate a period derivative of $\left(1.88\pm 0.08\right) \times 10^{-11}$~s~s$^{-1}$. MG05 also observed that the spectra can be well-fitted using a blackbody plus power-law model, and that the fits are significantly better than single component (blackbody or power-law) fits. On the other hand, TE08 fitted \textit{XMM--Newton} observations from 2000 to 2005 with a double blackbody model, and proposed that the cooler blackbody emission ($kT\sim 0.35~$keV) possibly arises from a significant portion of the magnetar surface. The authors stated the radius ($\sim 12$~km) of the cooler blackbody component as a lower limit to the magnetar radius. Their analysis also revealed a double-peaked pulse profile with no notable energy-dependence. 
\par \citet{durant2005} suggested a possible optical counterpart of \mgr$\,$ using archival \textit{Hubble Space Telescope} Wide Field Planetary Camera 2 (WFPC2) observations. Later, \citet{durant2008} imaged the same region using WFPC2 aiming to confirm the earlier detection although unfortunately, the source originally identified was not confirmed. 
\par In this paper, we present a detailed analysis of eleven observations of \mgr , spanning 2000 to 2016, carried out with the \textit{XMM-Newton} telescope. This includes the re-analysis of five older datasets (2000 to 2005; see MG05 and TE08) using the latest calibration files. We find that \mgr$\,$ has been steadily spinning down over the entire period of the observations, as expected from the magnetar model. We also calculate the spin-down rate $\dot{P}$ and pulse fraction of this magnetar. We carry out detailed spectral modeling of the source and provide an alternate prescription to explain the observed characteristics of its emission.
\par The paper is organized as follows: in Section~\ref{sec:obs}, we describe the observations that we analyzed and the data reduction process. In Section~\ref{sec:analysis}, we describe the timing and spectral analysis of the data. Section~\ref{sec:res} highlights our main results. Section~\ref{sec:discuss} presents a review of our results and their implications, and in Section~\ref{sec:summary}, we conclude with a brief summary.

\section{Observations and data reduction}
\label{sec:obs}
\textit{XMM-Newton} comprises of three X-ray cameras collectively known as European Photon Imaging Cameras (EPIC), of which two are MOS-CCD arrays \citep{turner2001} and the third is a PN-CCD \citep{struder2001}. Here, we have used archival data from the \textit{XMM-Newton} PN-CCD, which has a time resolution of 73.4~ms in the full frame imaging mode, for the timing analysis. The MOS data have a much poorer time resolution of 2.6~s and hence have been excluded from timing analysis. However, data from both types of CCDs have been employed for performing simultaneous spectral fits\footnote{In some of the observations, the source falls on one or more chip gaps (dead spaces due to detector edges) in the MOS1 and/or MOS2 exposures. Only those data have been utilized where the source is fully located on any one chip. See Table~\ref{tab:obs} for details.}. The observation date, MJD of the start time along with PN and MOS exposure time for the eleven datasets that we have analyzed are presented in Table~\ref{tab:obs}. Although \mgr$\,$ was not the main target of these observations, it was always within the EPIC field of view with an offset of $\sim 6'$ from the target in each observation (see Table~\ref{tab:obs}). The mean count rate and hardness (ratio of $1.3-8$~keV counts and $0.3-1.3$~keV counts) of each observation are also listed in Table~\ref{tab:obs}. The last column gives the identifier for each observation, which we will use to refer to them hereafter. All the PN observations were taken in the full frame mode (except observation A, which was in the extended full frame mode with a time resolution of 199~ms) with the medium optical blocking filter. 

\begin{table}
\setlength{\tabcolsep}{2.0pt}
\linespread{0.8}\selectfont\centering
\centering
	\caption{\textit{XMM-Newton} observation log of \mgr. The mean count rate as observed by the PN detector, and hardness of each observation are also listed.}
	\label{tab:obs}
	\begin{threeparttable}
	\scriptsize
	\bgroup
	\def\arraystretch{1.4}
	\resizebox{\columnwidth}{!}{
		\begin{tabular}{cccccccccc}
		\toprule\midrule
		\multirow{2}{*}{Obs ID} & \multirow{2}{*}{\begin{tabular}[c]{@{}c@{}}Offset\\ ($''$)\end{tabular}} & \multirow{2}{*}{\begin{tabular}[c]				{@{}c@{}}Start\\ date\end{tabular}} & \multirow{2}{*}{\begin{tabular}[c]{@{}c@{}}MJD\\ (days)\end{tabular}} & \multicolumn{3}{c}					{\begin{tabular}[c]{@{}c@{}}Exposure\\ time (ks)\end{tabular}} & \multirow{2}{*}{\begin{tabular}[c]{@{}c@{}}Count rate\\ (s$^{-1}$)					\end{tabular}} & \multirow{2}{*}{\begin{tabular}[c]{@{}c@{}}Hardness\\ (H/S)\end{tabular}} & \multirow{2}{*}{ID} \\\cline{5-7}
 &  &  &  & PN & MOS1 & MOS2 &  &  &  \\\hline
		0110000201 & 6.04 & Oct 17 2000 & 51834.7 & 15.9 & 19.7 & 19.7\tnote{*} & $0.194\pm 0.005$ & $0.556\pm 0.023$ & A  \\
		0018540101 & 6.05 & Nov 20 2001 & 52234.0 & 24.4 & 27.0\tnote{*} & 27.0\tnote{*} & $0.229\pm 0.005$ & $0.480\pm 0.024$ & B \\
		0304250401 & 6.05 & Nov 27 2005 & 53701.3 & 15.9 & 17.6 & 17.6\tnote{*} & $0.230\pm 0.005$ & $0.575\pm 0.020$ & C \\
		0304250501 & 6.05 & Nov 29 2005 & 53703.2 & 14.9 & 16.6 & 16.6\tnote{*} & $0.185\pm 0.006$ & $0.498\pm 0.033$ & D \\
		0304250601 & 6.05 & Dec 02 2005 & 53715.6 & 10.6 & 16.7 & 16.7 & $0.230\pm 0.005$ & $0.491\pm 0.027$ & E \\
		0780090301 & 6.07 & Sep 19 2016 & 57650.6 & 20.0 & 21.7 & 21.6 & $0.216\pm 0.005$ & $0.491\pm 0.023$ & F \\
		0780090401 & 6.07 & Sep 21 2016 & 57652.6 & 24.9 & 26.6 & 26.5 & $0.226\pm 0.004$ & $0.496\pm 0.022$ & G \\
		0780090501 & 6.07 & Oct 15 2016 & 57676.5 & 26.0 & 27.7\tnote{*} & 27.6\tnote{*} & $0.226\pm 0.003$ & $0.494\pm 0.020$ & H \\
		0780090601 & 6.07 & Oct 19 2016 & 57680.4 & 17.7 & 25.7\tnote{*} & 25.6\tnote{*} & $0.223\pm 0.005$ & $0.542\pm 0.022$ & I \\
		0780090701 & 6.07 & Oct 23 2016 & 57684.2 & 22.0 & 23.7\tnote{*} & 23.6\tnote{*} & $0.229\pm 0.004$ & $0.500\pm 0.021$ & J \\
		0780090901 & 6.07 & Dec 12 2016 & 57724.7 & 29.8 & 31.4\tnote{*} & 31.4\tnote{*} & $0.214\pm 0.004$ & $0.498\pm 0.021$ & K \\\hline
	\end{tabular}%
	}
	\egroup
	\begin{tablenotes}\scriptsize
	\item[*] Data could not be used for spectral analysis as part of the source falls in one of the chip gaps of the frame.
	\end{tablenotes}
	\end{threeparttable}
\end{table}

\par These eleven datasets (Table~\ref{tab:obs}) comprise all but two (observation IDs 0212282601 and 0780090801) available \textit{XMM-Newton} observations, which include this source within an offset of $< 10'$ (three more observations with higher offsets were not considered). ObsID~0212282601 was excluded as PN data was not available for this observation whereas the exposure time of ObsID~0780090801 was insufficient for a detailed analysis.

\par The data were reduced using the \textit{XMM-Newton} Science Analysis System (SAS version 18.0.0\footnote{https://www.cosmos.esa.int/web/xmm-newton/download-and-install-sas}) software package and the calibration files (CCF) released on October~2019. We generated calibrated PN and MOS event files using tasks \texttt{epchain} and \texttt{emchain} respectively, with the latest CALDB database. We applied barycentric correction to the event arrival times and then screened the data using standard filters to select good events in the range 0.3 to 10~keV. We defined the `source' as a circular region of radius $30''$ for PN data and $25''$ for MOS data, centered at RA = $01^h00^m43^s.14$, DEC = $-72^{\circ}11'33.8''$ \citep{lamb2002} and `background' as a nearby source-free region of $40''$ radius located on the same chip.

\par We checked for charged particle events in the energy range $12-15$~keV but did not find any significant signs of contamination. We extracted the source and background spectra for each observation using the task \texttt{evselect} and also generated the redistribution matrix file (RMF) and ancillary response file (ARF) for each spectrum using tasks \texttt{rmfgen} and \texttt{arfgen} respectively. The source and background light curves were binned at 146~ms, an integral multiple of the PN full frame time resolution of 73~ms\footnote{Observation A was made in the extended full frame mode with a time resolution of 199~ms, so we binned it at 398~ms.}, and extracted them using \texttt{evselect} task. The background-subtracted PN light curves were corrected for various instrumental effects such as vignetting, bad pixels, PSF variation and dead time using the task \texttt{epiclccorr}.

\section{Data analysis and modeling}
\label{sec:analysis}

\subsection{Timing Analysis}
\label{subsec:time}

First, the fractional root mean square variability ($F_{\mathrm{var}}$) of the background-subtracted light curve was calculated following the prescription of \citet{vaughan2003} as,
\begin{equation}
F_{\mathrm{var}} = \sqrt{\frac{S^2-\overline{\sigma^{\, 2}_{\mathrm{err}}}}{\bar{x}^2}}.
\label{eq:fvar}
\end{equation}
Here, $S^2$ is the variance, $\overline{\sigma_{\mathrm{err}}^{\, 2}}$ is the mean square error associated with each measurement and $\bar{x}$ is the mean count rate of the light curve. 

\par
\begin{figure}
  \centering
  \includegraphics[scale=0.4, trim = 0.4cm 0cm 0cm 0cm, clip = true]{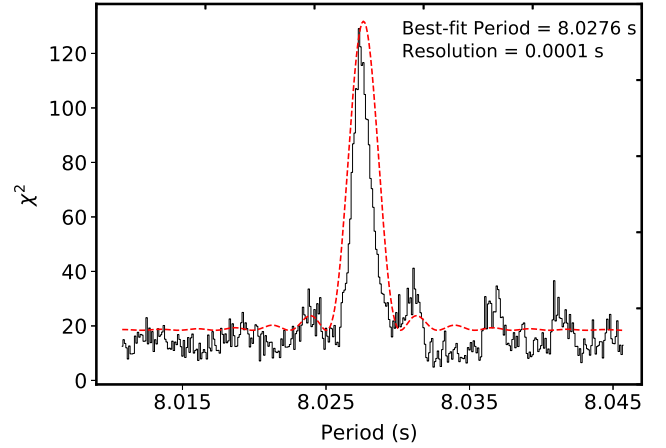}
  \caption{Result of period search using \textit{efsearch} for observation~G of \mgr. The dashed curve shows the fit to $\chi^2$ vs period obtained from \textit{efsearch} using the theoretical function (Eqn.~4) given in \citet{leahy1987}. The search resolution and best-fit period are indicated on the top right of the plot.}
\label{fig:efsearchfit}
\end{figure}

 We searched for periodicity in the binned time series using the epoch-folding task \textit{efsearch}\footnote{https://heasarc.gsfc.nasa.gov/xanadu/xronos\label{note1}} \citep{leahy1983} which is available as part of XRONOS package of HEASARC. \textit{efsearch} folds the data over a range of periods and searches for the maximum chi-square $(\chi^2)$ as a function of period. The task takes a light curve, epoch of observation, initial guess value for period, search resolution and range of periods to search as input. We used the period (P) of 8.0204~s and spin-down rate ($\dot{P}$) of $1.88\times 10^{-11}$~s~s$^{-1}$ quoted by MG05 as of January~28,~2004  as reference and calculated the expected spin periods at the epoch of each of our observations, which we then provided as inputs to \textit{efsearch} for initial period values. We assumed that the period does not change significantly over the span of a single observation and hence set the parameter \texttt{dpdot} to 0. The search resolution was set to 0.0001~s. \textit{efsearch} returns the $\chi^2$ as a function of period, which is then fitted with the theoretical function \citep{leahy1987} to obtain the best-fit period at each epoch. Figure~\ref{fig:efsearchfit} shows the theoretical fit (dashed line) to \textit{efsearch} output for observation~G. Statistical fluctuations make the chi-square peak returned by efsearch sensitive to binning, resolution and other factors and hence, a least-squares fitting of the peak provides a better estimate of the pulse period. Moreover, the theoretical function assumes the incoming signal to be perfectly sinusoidal. These factors contribute to the slight offset between the peak of the theoretical best fit and the chi-square peak as seen in Figure~\ref{fig:efsearchfit}.

\par To estimate the uncertainties in the best-fit periods, Monte Carlo simulations were performed to generate photon arrival time series (assuming the incoming signal is sinusoidal) with counting statistics, which were then epoch-folded using \textit{efsearch} and the $\chi^2$ peaks were fitted following the prescription of \citet{leahy1987}. Corresponding to each observation, 50~simulations were performed and their standard deviation was quoted as the uncertainty in the period measurement.

\par We then computed the spin-down rate of the source by performing a linear regression on the obtained best-fit periods. We also included the periods 8.0188~s, 8.0204~s (corresponding to May~15~2001 and January~28~2004 respectively as quoted by MG05) and 8.0215~s (on March~27~2005 quoted by TE08) in our fits.

\par Further, to investigate the timing behaviour of the source and its energy-dependence, we extracted source and background light curves in the soft ($0.3-1.3$~keV) and hard ($1.3-8$~keV) band. The boundary between the two energy bands was chosen so as to ensure that most of the cooler blackbody photons are included in the soft band\footnote{This was motivated by the double blackbody fits of TE08, where the cooler blackbody has $kT\sim 0.3$~keV.}. We verify that this is indeed true by fitting the spectra with double blackbodies and explicitly calculating the flux of each blackbody in both bands (see Section~\ref{subsec:spec}). The upper limit for the hard band is set keeping in mind the poor statistics beyond 8~keV.

\par We produced background-subtracted folded light curves (task \textit{efold}\textsuperscript{\ref{note1}}) in both energy bands defined above, using best-fit pulse periods obtained for the respective epochs of observation. A binning of 16 per phase in the folded light curves adequately demonstrates the emission modulation. \textit{efold} returns the count rate in each bin, along with their $1\sigma$ uncertainties. We also calculated the hardness ratio defined by $(H-S)/(H+S)$ in TE08, where $H$ and $S$ are the count rates in hard and soft band respectively, and the associated $1\sigma$ uncertainties. We then computed the pulse fraction ($PF$) in both energy bands using the relation \citep[e.g.][]{ho2007},
\begin{equation}
PF=\frac{C_{\mathrm{max}}-C_{\mathrm{min}}}{C_{\mathrm{max}}+C_{\mathrm{min}}},
\end{equation}
where $C_{\mathrm{max}}$ and $C_{\mathrm{min}}$ are the maximum and minimum count rates per bin, respectively, for a given peak during a pulse cycle. The $1\sigma$ uncertainties in the pulse fraction were calculated by propagating the uncertainties in $C_{\mathrm{max}}$ and $C_{\mathrm{min}}$ returned by \textit{efold}.

\subsection{Spectral Analysis}
\label{subsec:spec}

The X-ray spectra generated by SAS were analyzed using XSPEC\footnote{https://heasarc.gsfc.nasa.gov/xanadu/xspec/} version 12.10.1. The phase-averaged source spectra were re-binned to have a minimum of 30 counts per bin. Wherever MOS spectra were available (see Table~\ref{tab:obs}), the PN and MOS spectra were simultaneously fitted with corresponding parameters tied to each other and the normalizations left to vary freely to account for the cross-calibrations between the instruments.

\par We attempted fitting several single and double component models to the spectra in the 0.3 to 10~keV energy range. Two-component models such as double blackbody and blackbody plus power-law produced acceptable fits ($\chi^2_{\mathrm{red}}$ in the range $0.73-1.33$ and $0.83-1.39$ respectively) to the spectra of all observations. For the double blackbody fits, we calculated the contribution of PN flux of the hotter blackbody (average \textit{kT}$=0.57$~keV) in the soft band ($0.3-1.3$~keV) to be only $19.7\%$ on average. On the other hand, the mean PN fluxes of the hotter and cooler blackbodies in the hard band ($1.3-8$~keV) were 66\% and 34\% respectively. This would imply that in the soft band, the cooler blackbody majorly contributes in producing the double-peaked profile whereas in the hard band, the hotter blackbody dominates. The observed similarity of pulse profiles in the two energy bands can be much more simply and conveniently explained by considering a single component, arising from a single physical process, which is responsible for emission throughout the energy range (also see Section~\ref{sec:discuss}). Therefore, we did not pursue here two-component models further.

\par We fit the following single component models to the spectra: blackbody (\texttt{bbodyrad}), power-law (\texttt{powerlaw}), power-law with high energy exponential cutoff (\texttt{cutoffpl}) and thermally Comptonized continuum (\texttt{nthcomp}, \citealt{zdziarski1996,zycki1999}), each modified by photoelectric absorption\footnote{We also attempted fitting the RCS model \citep{rea2008} but for many of the observations, this model was unable to constrain the parameters.}. We used the Tuebinger-Boulder ISM absorption model (\texttt{tbabs} in XSPEC) of \citet{wilms2000} to estimate the hydrogen column density ($N_H$). For each fit, we allowed the column density to vary between $(0.06-0.5)\times 10^{22}$~cm$^{-2}$. The lower limit is the average $N_H$ at the coordinates of this source taken from \citet{dickey1990}. However, more recent surveys (e.g. \citealt{kalberla2005, bekhti2016}) give $N_H$ estimates which are an order of magnitude higher than the previously adopted value. The former quotes a value of $3.1\times 10^{21}$~cm$^{-2}$ whereas the latter survey gives a value of $4.5\times 10^{21}$~cm$^{-2}$, which dictated the upper limit of $N_H$ in our fits. 

\par On fitting the blackbody model to the spectra, the column density value hits the allowed lower limit of $6\times 10^{20}$~cm$^{-2}$ for each observation. Despite constraining the blackbody temperature well, this model produces the highest chi-square values among the four models ($\chi^2_{\mathrm{red}} > 1.3$ for most observations; see Table~\ref{bbr_fit}). This indicates that the fits are barely acceptable. Hence, we do not proceed with this model further. 

\par A power-law model produces better fits compared to the blackbody model for all observations (Table~\ref{pow_fit}), although $\chi^2_{\mathrm{red}}$ values are still quite high ($\chi^2_{\mathrm{red}} > 1.1$ for all observations except B and J). The column density values in this case are well-constrained between ($0.37-0.46)\times 10^{22}$~cm$^{-2}$, in agreement with both \citet{kalberla2005} and \citet{bekhti2016}. Although this model constrains the power-law index well, the cutoff power-law model as well as the Comptonization model (Tables~\ref{cutoffpl_fit} and \ref{nthcomp_fit} respectively) significantly improve the fits for each observation and thus the power-law model is rejected in favour of the other two.

\par For the cutoff power-law model, the photon index as well as the high-energy cutoff show a large variation and significant uncertainties. The $N_H$ for this model varies in the range ($0.1-0.3)\times 10^{22}$~cm$^{-2}$. Apart from the constraints on $N_H$, all other parameters are allowed to vary freely in the spectral fits with the blackbody, power-law and cutoff power-law models.
\par The lower energy statistics are not sufficient to constrain the seed photon temperature (assumed to have a blackbody distribution) for the thermal Comptonization model, and hence we fix it at 0.2~keV, which is the average best-fit value if this parameter were to be left free. Based purely on goodness-of-fit considerations, both cutoff power-law model and the Comptonization model perform equally well, with the latter fits only slightly better than the former in most cases. However, despite the cutoff power-law model giving good fits to the data, five of the observations have photon indices less than 1 which corresponds to negative values of the Compton $y$ parameter (\citealt{lavagetto2008}, \citealt{agrawal2020}). This makes the model an incorrect description of Comptonized emission and is hence deemed unfit on physical grounds.

\par
\begin{figure}
  \centering
  \includegraphics[scale=0.4, trim = 0cm 0.8cm 0.5cm 1.3cm, clip = true]{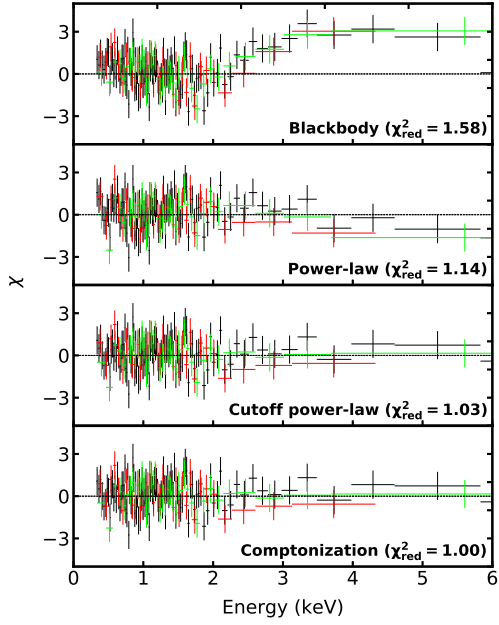}
  \caption{Residuals of the four models -- (top to bottom) blackbody, power-law, cutoff power-law and thermal Comptonization -- fitted to the observation~G. The black, red and green points respectively represent the PN, MOS1 and MOS2 data. The $\chi^2_{\mathrm{red}}$ values are mentioned in the bottom right corner of each panel.}
\label{fig:residuals}
\end{figure}

 Figure~\ref{fig:residuals} shows the residuals left after fitting the spectra with the four models, considering the example of observation~G.
The quality of the fits improves from the top to the bottom panel, with the $\chi^2_{red}$ reducing from 1.58 for the blackbody model to 1.00 for the Comptonization model. In the next section, we present the results of our timing and spectral analyses.  

\section{Results}
\label{sec:res}  

\subsection{Timing Properties}
\label{subsec:res_time}

\begin{figure*}
\centering
	\includegraphics[width = \textwidth, scale = 0.7, trim = 3.5cm 0cm 4cm 0cm, clip = true]{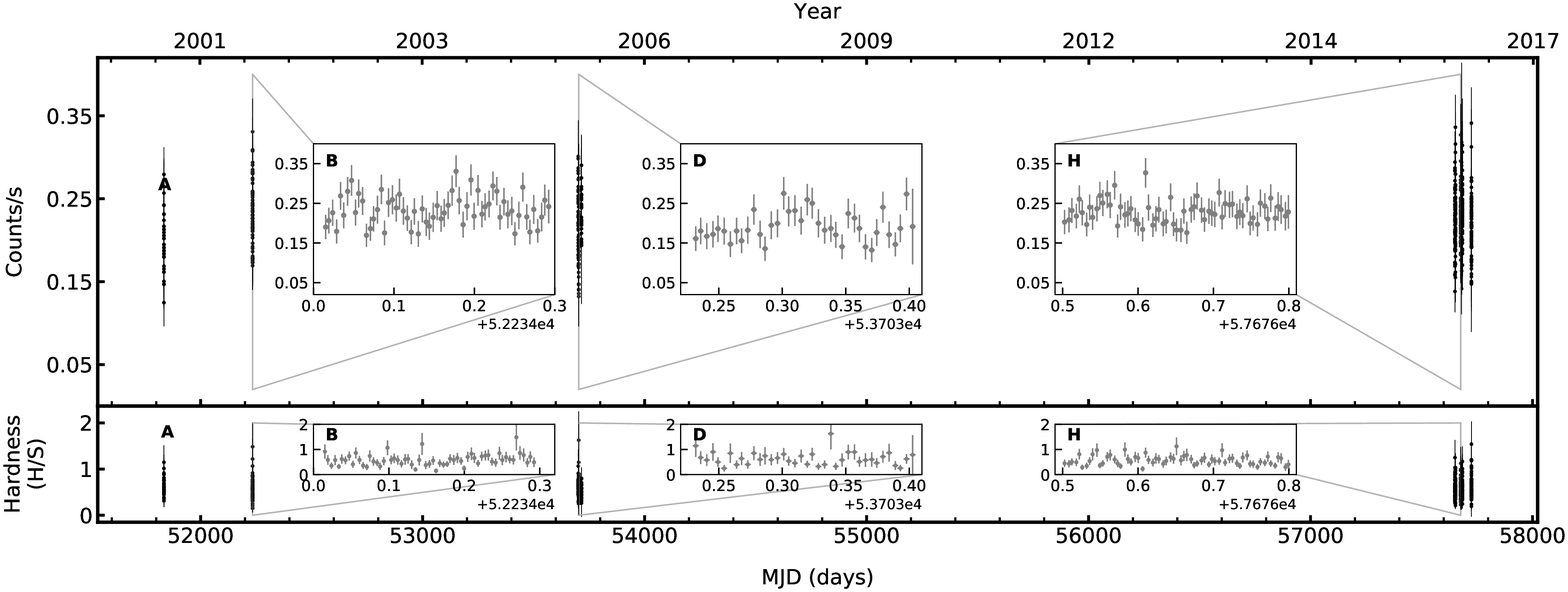}
    \caption{Background subtracted light curve (top) and hardness (bottom, see text for details) over the entire span of observations. The inset plots show them zoomed in for (left to right) observations~B, D and H. Each black dot (grey for the insets) corresponds to data binned at 400~s. Observation~A, carried out on October~17~2000, corresponds to MJD~$51834.7$.}
    \label{fig:ext-lc}
\end{figure*}

Figure~\ref{fig:ext-lc} shows the variation of count rate and hardness (defined as the ratio of $1.3-8$~keV counts and $0.3-1.3$~keV counts) of \mgr$\,$ over the entire period of observations. Insets are provided for observations~B, D and H to demonstrate the variabilities within the observation period. The light curve clearly demonstrates that this is a steady source. Indeed, the fractional variability in the entire light curve (see Equation~\ref{eq:fvar}) is just $4.69\%$, with mean count rate varying in the narrow range $0.19-0.23$~counts~s$^{-1}$. The mean hardness lies in the range $0.48-0.58$ with a standard deviation of just 0.03.

\begin{figure}
\centering
	\includegraphics[scale = 0.5]{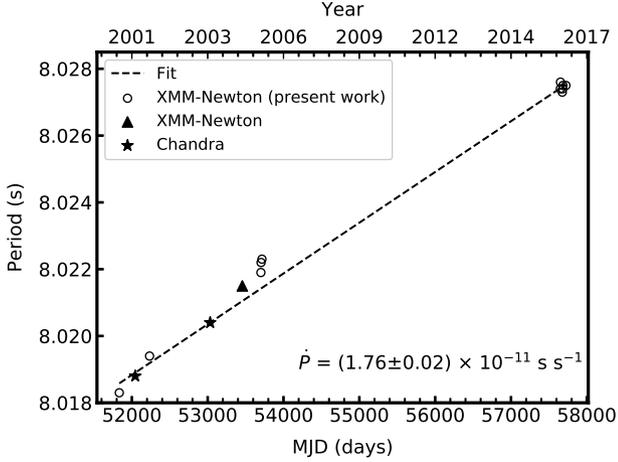}
    \caption{Spin period of \mgr$\,$ calculated at different epochs of observation from 2000 to 2016. Circles mark the periods computed in this work whereas stars and the triangle are values taken from MG05 and TE08 respectively. The error bars are smaller than the symbol size. The dashed line is the best fit straight line passing through all the fourteen points. The spin-down rate computed from the fit is $1.76\times 10^{-11}$~s~s$^{-1}$. See text for details.}
    \label{fig:pdot}
\end{figure}

As explained in Section~\ref{subsec:time}, we determined the pulse period of the source at each epoch of observation. In Table~\ref{tab:timing} (column~2), we mention the pulse-periods obtained with \textit{efsearch}. One can see a clear trend of increasing spin periods from observation A to K. Figure~\ref{fig:pdot} plots the evolution of the spin periods against their date of observation. Two \textit{Chandra} and one \textit{XMM-Newton} data points (taken from MG05 and TE08 respectively; see Section~\ref{subsec:time}) are also included. The slope of the best-fit straight line, $(1.76\pm 0.02) \times 10^{-11}$~s~s$^{-1}$, gives the period derivative of this source over the span of observations analyzed. The source is seen to steadily spin-down from 8.0183(2)~s in October~2000 to 8.0275(1)~s in December~2016. 

\begin{table}
\centering
\caption{Spin-period of \mgr$\,$ and its pulse fraction in $0.3-1.3$ and $1.3-8$~keV bands. The numbers in parentheses indicates the uncertainty in the last digit. For each energy band, the pulse fraction of both peaks (designated `broad' and `narrow') are separately calculated. The last row quotes the percentage RMS fluctuation in the pulse fractions.}
\label{tab:timing}
\begin{threeparttable}
\resizebox{\columnwidth}{!}{
	\begin{tabular}{cccccc}
	\toprule \midrule
	\multirow{3}{*}{Obs} & \multirow{3}{*}{P (s)} & \multicolumn{4}{c}{Pulse fraction (\%)} \\ 
	 &  & \multicolumn{2}{c}{$0.3-1.3$ keV} & \multicolumn{2}{c}{$1.3-8.0$ keV} \\ \cline{3-6}
	 &  & Broad & Narrow & Broad & Narrow \\\hline
	A & 8.0183(2) & $38.13\pm 8.97$ & $34.23\pm 9.23$ & $41.49\pm 13.47$ & $39.64\pm 13.76$ \\
	B & 8.0194(1) & $39.57\pm 6.78$ & $25.29\pm 7.28$ & $45.28\pm 11.10$ & $34.56\pm 11.83$ \\
	C & 8.0219(3) & $35.05\pm 9.03$ & $25.59\pm 9.40$ & $36.74\pm 11.12$ & $26.41\pm 11.65$ \\
	D & 8.0222(2) & $38.46\pm 11.24$ & $27.12\pm 11.67$ & $50.78\pm 15.54$ & $34.08\pm 17.47$ \\
	E & 8.0223(4) & $31.39\pm 9.74$ & $25.22\pm 10.09$ & $56.55\pm 15.85$ & $41.41\pm 17.97$ \\
	F & 8.0274(1) & $47.45\pm 8.25$ & $41.19\pm 8.66$ & $41.30\pm 9.93$ & $25.52\pm 10.95$ \\
	G & 8.0276(1) & $32.92\pm 6.54$ & $19.39\pm 7.03$ & $37.14\pm 9.12$ & $31.11\pm 9.33$ \\
	H & 8.0273(1) & $33.80\pm 6.73$ & $32.63\pm 6.75$ & $43.85\pm 9.59$ & $31.70\pm 10.24$ \\
	I & 8.0274(2) & $46.53\pm 8.59$ & $34.54\pm 9.05$ & $40.42\pm 11.74$ & $34.32\pm 12.34$ \\
	J & 8.0275(2) & $28.56\pm 6.84$ & $19.72\pm 7.17$ & $37.69\pm 9.60$ & $17.97\pm 10.72$ \\
	K & 8.0275(1) & $32.25\pm 6.76$ & $23.37\pm 7.04$ & $63.03\pm 14.69$ & $57.83\pm 16.02$ \\\hline
	\multicolumn{2}{c}{RMS variation (\%)} & 15.68 & 23.14 & 18.05 & 28.76\\\hline
	\end{tabular}
}
\end{threeparttable}
\end{table}

\par
\begin{figure*}
\centering
\begin{subfigure}{\columnwidth}
\includegraphics[width=\columnwidth, trim = 0cm 1cm 1.6cm 2cm, clip=true]{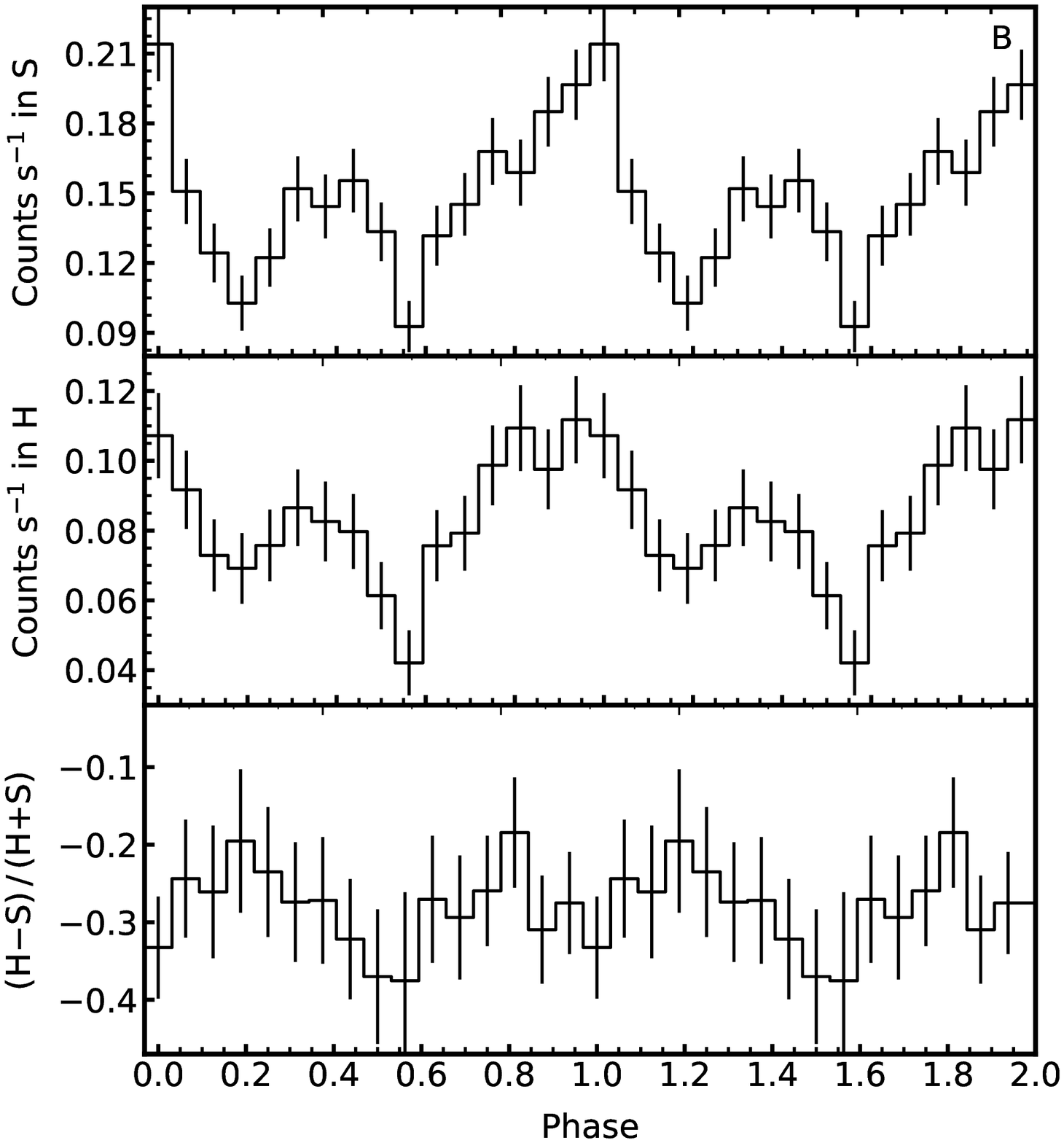}
\caption{}
\end{subfigure}\hfill
\begin{subfigure}{\columnwidth}
\includegraphics[width=\columnwidth, trim = 0cm 1cm 1.6cm 2cm, clip=true]{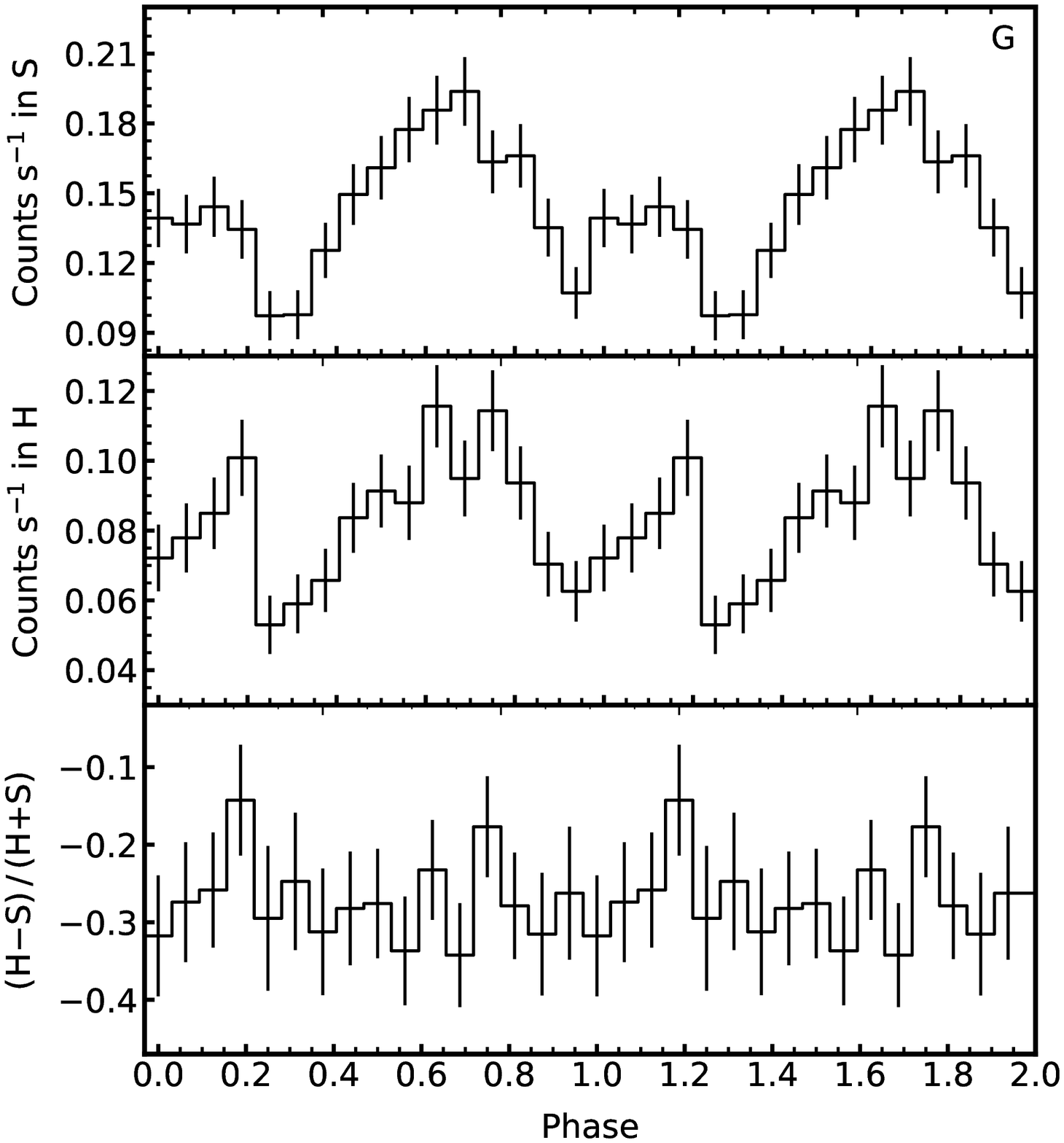}
\caption{}
\end{subfigure}\hfill
\caption{Folded pulse profiles and hardness ratio defined by $(H-S)/(H+S)$ for observation (a) B and (b) G. $H$ and $S$ are count rates in $1.3-8$~keV and $0.3-1.3$~keV respectively. In each figure, top panel shows the count rate in soft band, middle panel shows the count rate in hard band and bottom panel plots the hardness ratio. $1\sigma$ error bars are also plotted. See text for details.}
\label{fig:folded-lc}
\end{figure*}

 Further, we produced background-subtracted folded light curves of the source in both energy bands ($0.3-1.3$~keV and $1.3-8$~keV) to investigate the energy-dependence of the emission. We find that both soft and hard emission is characterized by clear double-peaked profiles with one broad and one narrow peak. Moreover, the modulation in the two bands is `synchronous' with each other -- they rise and fall at the same phases, with similar peak widths in both bands. Figure~\ref{fig:folded-lc} shows the folded light curves for observations B and G in the two energy bands, along with the corresponding hardness ratios (defined in Section~\ref{subsec:time}). In a few datasets, although the profile in the hard band is not as well-defined as that of the soft band owing to poor statistics, the synchronization of pulse profiles at the two energies is unmistakable. The mean hardness ratio of the observations is $-0.26$, with an average standard deviation of $0.02$. Thus, the hardness ratio remains roughly constant over the span of each observation, which further highlights the fact that the properties of emission in the soft band relative to the hard band does not significantly change with the rotational phase of the magnetar. Moreover, the average hardness ratio does not fluctuate significantly with time over the entire span of observations, varying in the narrow range of $-0.29$ to $-0.23$.

\par In Table~\ref{tab:timing}, we summarize the pulse fraction of the source in the two energy bands, along with their $1\sigma$ error bars. As mentioned above, the pulse profile is double-peaked and the label of `broad' and `narrow' is quite unambiguous. The pulse fractions in the soft band for the broad and narrow peaks lie in the range ($29-47$)\% and ($19-41$)\% respectively. In the hard band, the pulse fractions lie between ($18-58$)\% for the narrow peak and between ($37-63$)\% for the broad peak. It is to be noted that the pulse fractions (RMS variation 16 to 29$\%$, see Table~\ref{tab:timing}) are consistent with no temporal evolution.

\subsection{Spectral Properties}
\label{subsec:res_spec} 

As mentioned in Section~\ref{subsec:res_time}, the pulse profile in both energy bands are very similar and a single component spectral model provides a simple explanation for this similarity. We fit four single-component models to the phase-averaged PN and MOS spectra of each observation, as described in Section~\ref{subsec:spec}. The spectral parameters for the four models are presented in Tables~\ref{bbr_fit} to \ref{nthcomp_fit}. 

\par The blackbody temperatures obtained with the blackbody model are well-constrained in the range 0.34 to 0.36~keV (see Table~\ref{bbr_fit}). For a distance of 60~kpc to the SMC \citep{keller2006}, the range of blackbody radii is $9-10$~km, signifying thermal emission from a significant portion of the magnetar surface. Using a power-law model, the photon index has a narrow range of variation between 3.3 and 3.7 as can be seen from Table~\ref{pow_fit}. However, both cutoff power-law and Comptonization model (Tables~\ref{cutoffpl_fit} and \ref{nthcomp_fit}) prove superior compared to these two models in terms of their $\chi^2_{\mathrm{red}}$. Cutoff power-law model produces fits with photon indices varying from $0.59$ to $1.99$ and cutoff energy in the range $0.8-1.5$~keV. These two quantities are not well-constrained for most of the observations. Moreover, a photon index less than 1 is unphysical for describing Comptonized emission, as is the case of five of the observations. As explained in Section~\ref{subsec:spec}, we proceed with the thermal Comptonization model and reject the other three models. 

\par
\begin{figure}
\centering
	\includegraphics[scale=0.5, trim=0cm 1.45cm 0cm 0cm, clip=true, angle=-90]{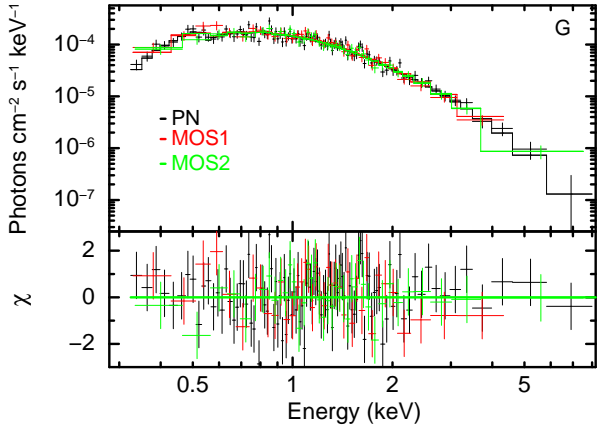}
    \caption{Phase-averaged spectral fit to observation G using thermal Comptonization model. The top and bottom panels show the unfolded spectra and residuals (in units of $\sigma$) respectively.}
    \label{fig:specfit}
\end{figure}

 Using the Comptonization model, we obtain the temperature of electron plasma in the range $0.54-0.87$~keV. In most of the cases, the electron temperature is well-constrained with uncertainties $<30\%$. The photon index is seen to vary between 2.03 to 2.73 with $1\sigma$ uncertainties $<15\%$. We are able to constrain $N_H$ in the narrow range from $(0.8-1.5)\times 10^{21}$~cm$^{-2}$ for all the observations. The unabsorbed X-ray ($0.3-10$~keV) flux of the source (averaged over PN and MOS) using this model varies between $(4.8-5.7)\times 10^{-13}$~erg~cm$^{-2}$~s$^{-1}$, which translates to source X-ray luminosity in the range $(2.1-2.5)\times 10^{35}$~erg~s$^{-1}$ for an assumed distance of 60~kpc. We do not observe any apparent correlation between the luminosity and hardness of emission, with the latter remaining roughly constant over the observed luminosity range. We also examined the pulse fraction of the broad and narrow peaks in the two bands as a function of luminosity but do not find any noteworthy trends. In Figure~\ref{fig:specfit}, we show the spectral fit of observation~G with the thermal Comptonization model, along with the residuals. 

\begin{table}
	\centering
	\caption{Best fit spectral parameters using blackbody model. $kT_{BB}$ is the blackbody temperature of the fit in keV. $N_H$ is in units of 10$^{22}$~cm$^{-2}$. Uncertainties quoted are $1\sigma$ error bars.}
	\label{bbr_fit}
	\begin{threeparttable}
	\scriptsize
	\bgroup
	\setlength{\tabcolsep}{2pt}
	\def\arraystretch{1.4}
	\begin{tabular}{ccccccc}
		\toprule \midrule		
		\multirow{2}{*}{Obs} & \multirow{2}{*}{$N_H$\tnote{*}} & \multirow{2}{*}{$kT_{BB}$} & \multicolumn{3}{c}{Norm} & \multirow{2}{*}					{\begin{tabular}[c]{@{}c@{}}$\chi^2_{\mathrm{red}}$ \\ ($\chi^2$/DOF)\end{tabular}} \\\cline{4-6}
		 &  &  & PN & MOS1 & MOS2 &  	\\	
		\hline
		A & 0.06 & $0.35^{+0.01}_{-0.01}$ & $2.15^{+0.23}_{-0.10}$ & $2.73^{+0.31}_{-0.15}$ & ... & 1.29 (120.26/93)  \\
		B & 0.06 & $0.35^{+0.01}_{-0.01}$ & $2.82^{+0.14}_{-0.24}$ & ... & ... & 1.30 (132.13/102)  \\
		C & 0.06 & $0.35^{+0.01}_{-0.01}$ & $2.61^{+0.19}_{-0.19}$ & $2.57^{+0.22}_{-0.23}$ & ... & 1.71 (137.04/80)  \\
		D & 0.06 & $0.35^{+0.01}_{-0.01}$ & $2.12^{+0.20}_{-0.19}$ & $2.48^{+0.26}_{-0.24}$ & ... & 1.81 (104.71/58) \\
		E & 0.06 & $0.35^{+0.01}_{-0.01}$ & $2.69^{+0.16}_{-0.26}$ & $2.22^{+0.15}_{-0.24}$ & $2.19^{+0.15}_{-0.23}$ & 1.67 (131.98/79) \\
		F & 0.06 & $0.36^{+0.01}_{-0.01}$ & $2.33^{+0.17}_{-0.10}$ & $2.32^{+0.19}_{-0.12}$ & $2.39^{+0.20}_{-0.13}$ & 1.82 (261.56/144) \\
		G & 0.06 & $0.35^{+0.01}_{-0.01}$ & $2.62^{+0.13}_{-0.12}$ & $2.69^{+0.16}_{-0.15}$ & $2.60^{+0.16}_{-0.14}$ & 1.58 (282.76/179) \\
		H & 0.06 & $0.34^{+0.01}_{-0.01}$ & $2.88^{+0.18}_{-0.17}$ & ... & ... & 1.56 (165.28/106) \\
		I & 0.06 & $0.36^{+0.01}_{-0.01}$ & $2.44^{+0.16}_{-0.21}$ & ... & ... & 1.27 (100.14/79) \\
		J & 0.06 & $0.34^{+0.01}_{-0.01}$ & $3.03^{+0.15}_{-0.27}$ & ... & ... & 1.25 (117.31/94) \\
		K & 0.06 & $0.36^{+0.01}_{-0.01}$ & $2.43^{+0.15}_{-0.14}$ & ... & ... & 1.47 (145.80/99) \\
		\hline
	\end{tabular}%
	\egroup
\begin{tablenotes}\scriptsize
\item[*] For every fit, the value of hydrogen column density hit the lower limit ($0.06\times 10^{22}$~cm$^{-2}$) of the allowed range of values.
\end{tablenotes}
\end{threeparttable}
\end{table}

\begin{table}
	\centering
	\caption{Best fit spectral parameters for power-law model. $\Gamma$ is the photon index of the fit. Norm is in units of $10^{-4}$~photons~keV$^{-1}$~cm$^{-2}$~s$^{-1}$ at 1~keV. $N_H$ is in units of 10$^{22}$~cm$^{-2}$. Uncertainties quoted are $1\sigma$ error bars.}
	\label{pow_fit}
	\begin{threeparttable}
	\scriptsize
	\bgroup
	\setlength{\tabcolsep}{2pt}
	\def\arraystretch{1.4}
	\begin{tabular}{ccccccc}
		\toprule \midrule
		\multirow{2}{*}{Obs} & \multirow{2}{*}{$N_H$} & \multirow{2}{*}{$\Gamma$} & \multicolumn{3}{c}{Norm} & \multirow{2}{*}					{\begin{tabular}[c]{@{}c@{}}$\chi^2_{\mathrm{red}}$ \\ ($\chi^2$/DOF)\end{tabular}} \\\cline{4-6}
		 &  &  & PN & MOS1 & MOS2 &  	\\
		\hline
		A & $0.44^{+0.02}_{-0.02}$ & $3.62^{+0.11}_{-0.10}$ & $3.94^{+0.31}_{-0.28}$ & $4.96^{+0.42}_{-0.38}$ & ... & 1.21 (112.70/93) \\
		B & $0.40^{+0.02}_{-0.02}$ & $3.38^{+0.10}_{-0.10}$ & $4.15^{+0.30}_{-0.27}$ & ... & ... & 0.90 (92.29/102) \\
		C & $0.39^{+0.02}_{-0.02}$ & $3.33^{+0.10}_{-0.10}$ & $3.95^{+0.32}_{-0.29}$ & $3.89^{+0.33}_{-0.30}$ & ... & 1.29 (103.53/80) \\
		D & $0.37^{+0.03}_{-0.03}$ & $3.31^{+0.12}_{-0.12}$ & $3.10^{+0.30}_{-0.27}$ & $3.75^{+0.35}_{-0.32}$ & ... & 1.21 (70.35/58) \\
		E & $0.44^{+0.03}_{-0.02}$ & $3.56^{+0.11}_{-0.11}$ & $4.66^{+0.41}_{-0.37}$ & $3.81^{+0.35}_{-0.32}$ & $3.85^{+0.34}_{-0.31}$ & 1.27 (100.18/79) \\
		F & $0.43^{+0.02}_{-0.02}$ & $3.56^{+0.08}_{-0.07}$ & $4.32^{+0.26}_{-0.24}$ & $4.38^{+0.28}_{-0.26}$ & $4.47^{+0.29}_{-0.27}$ & 1.60 (230.71/144) \\
		G & $0.43^{+0.02}_{-0.02}$ & $3.53^{+0.07}_{-0.07}$ & $4.39^{+0.24}_{-0.22}$ & $4.53^{+0.26}_{-0.25}$ & $4.38^{+0.26}_{-0.24}$ & 1.14 (203.29/179) \\
		H & $0.39^{+0.02}_{-0.02}$ & $3.43^{+0.09}_{-0.09}$ & $3.95^{+0.27}_{-0.25}$ & ... & ... & 1.22 (129.57/106) \\
		I & $0.44^{+0.03}_{-0.03}$ & $3.53^{+0.12}_{-0.12}$ & $4.51^{+0.40}_{-0.36}$ & ... & ... & 1.17 (92.41/79) \\
		J & $0.39^{+0.02}_{-0.02}$ & $3.45^{+0.10}_{-0.09}$ & $4.03^{+0.29}_{-0.27}$ & ... & ... & 1.07 (100.36/94) \\
		K & $0.46^{+0.02}_{-0.02}$ & $3.68^{+0.10}_{-0.09}$ & $4.76^{+0.33}_{-0.30}$ & ... & ... & 1.40 (138.44/99) \\
		\hline
	\end{tabular}%
	\egroup
	\end{threeparttable}
\end{table}

\begin{table}
	\centering
	\caption{Best fit spectral parameters for cutoff power-law model. $\Gamma$ is the photon index and $E_{\mathrm{cut}}$ is the exponential cutoff energy of the fit in keV. Norm is in units of $10^{-4}$~photons~keV$^{-1}$~cm$^{-2}$~s$^{-1}$ at 1~keV. $N_H$ is in units of 10$^{22}$~cm$^{-2}$. Uncertainties quoted are $1\sigma$ error bars.}
	\label{cutoffpl_fit}
	\begin{threeparttable}
	\scriptsize
	\bgroup
	\setlength{\tabcolsep}{2pt}
	\def\arraystretch{1.4}
	\begin{tabular}{ccccccccc}
		\toprule \midrule
		\multirow{2}{*}{Obs} & \multirow{2}{*}{$N_H$} & \multirow{2}{*}{$\Gamma$} & \multirow{2}{*}{$E_{\mathrm{cut}}$} & 							\multicolumn{3}{c}{Norm} & \multirow{2}{*}{\begin{tabular}[c]{@{}c@{}}$\chi^2_{\mathrm{red}}$ \\ ($\chi^2$/DOF)\end{tabular}} \\					\cline{5-7}
		 &  &  &  & PN & MOS1 & MOS2 &  	\\
		\hline
		A & $0.16^{+0.06}_{-0.07}$ & $0.59^{+0.61}_{-0.88}$ & $0.76^{+0.20}_{-0.18}$ & $6.82^{+1.84}_{-0.98}$ & $8.67^{+2.37}_{-1.28}$ & ... & 1.00 (91.68/92)\\
		B & $0.23^{+0.05}_{-0.03}$ & $1.54^{+0.53}_{-0.56}$ & $1.29^{+0.56}_{-0.32}$ & $5.55^{+0.90}_{-0.68}$ & ... & ... & 0.78 (78.52/101) \\
		C & $0.23^{+0.06}_{-0.06}$ & $1.62^{+0.62}_{-0.68}$ & $1.37^{+0.83}_{-0.41}$ & $5.26^{+1.03}_{-0.73}$ & $5.13^{+1.02}_{-0.72}$ & ... & 1.20 (94.60/79) \\
		D & $0.20^{+0.08}_{-0.08}$ & $1.48^{+0.82}_{-0.88}$ & $1.27^{+1.07}_{-0.43}$ & $4.31^{+1.10}_{-0.76}$ & $5.20^{+1.33}_{-0.50}$ & ... & 1.13 (64.39/57) \\
		E & $0.29^{+0.07}_{-0.07}$ & $1.99^{+0.70}_{-0.71}$ & $1.46^{+1.16}_{-0.47}$ & $6.18^{+1.22}_{-0.92}$ & $5.12^{+1.04}_{-0.78}$ & $5.11^{+1.01}_{-0.76}$ & 1.21 (72.33/60) \\
		F & $0.16^{+0.04}_{-0.04}$ & $0.69^{+0.49}_{-0.50}$ & $0.83^{+0.18}_{-0.13}$ & $6.75^{+0.93}_{-0.74}$ & $6.88^{+0.96}_{-0.73}$ & $6.97^{+0.98}_{-0.79}$ & 1.33 (190.53/143) \\
		G & $0.25^{+0.04}_{-0.04}$ & $1.74^{+0.42}_{-0.44}$ & $1.34^{+0.43}_{-0.27}$ & $5.72^{+0.63}_{-0.52}$ & $5.88^{0.66}_{-0.55}$ & $5.72^{+0.65}_{-0.54}$ & 1.03 (182.56/178) \\
		H & $0.16^{+0.06}_{-0.05}$ & $0.84^{+0.72}_{-0.55}$ & $0.88^{+0.37}_{-0.17}$ & $6.42^{+1.16}_{-1.11}$ & ... & ... & 1.05 (110.24/105) \\
		I & $0.20^{+0.07}_{-0.07}$ & $0.94^{+0.76}_{-0.78}$ & $0.89^{+0.39}_{-0.22}$ & $7.25^{+1.80}_{-1.26}$ & ... & ... & 1.02 (79.21/78) \\
		J & $0.14^{+0.05}_{-0.06}$ & $0.59^{+0.66}_{-0.71}$ & $0.79^{+0.26}_{-0.17}$ & $7.26^{+1.84}_{-1.25}$ & ... & ... & 0.84 (77.73/93) \\
		K & $0.22^{+0.05}_{-0.06}$ & $1.10^{+0.53}_{-0.73}$ & $0.91^{+0.25}_{-0.21}$ & $7.36^{+1.60}_{-0.92}$ & ... & ... & 1.18 (116.06/98) \\
		\hline
	\end{tabular}%
	\egroup
	\end{threeparttable}
\end{table}

\begin{table}
	\centering
	\caption{Best fit spectral parameters for thermal Comptonization model. $\Gamma$ is the photon index, $kT_e$ is the electron temperature (in keV) of the fit. Seed photon temperature ($kT_{BB}$) is fixed at 0.2~keV for all the fits. $N_H$ is in units of 10$^{22}$~cm$^{-2}$. The unabsorbed flux ($0.3-10$~keV, PN and MOS averaged), optical depth of the electron plasma and Compton $y$ parameter are also shown. Uncertainties quoted are $1\sigma$ error bars.}
	\label{nthcomp_fit}
	\begin{threeparttable}
	\scriptsize
	\bgroup
	\setlength{\tabcolsep}{2pt}
	\def\arraystretch{1.4}
	\begin{tabular}{cccccccc}
		\toprule \midrule
		
		\multirow{2}{*}{Obs} & \multirow{2}{*}{$N_H$} & \multirow{2}{*}{$\Gamma$} & \multirow{2}{*}{$kT_e$} & \multirow{2}{*}{\begin{tabular}				[c]{@{}c@{}}$\chi^2_{\mathrm{red}}$ \\ ($\chi^2$/DOF)\end{tabular}} & \multicolumn{3}{c}{Derived parameters} \\ \cline{6-8} 
 		 & & & & & Flux\tnote{*} & $\tau$ & $y$ \\ 					
		\hline		
		A & $0.10^{+0.02}_{-0.03}$ & $2.08^{+0.28}_{-0.27}$ & $0.54^{+0.09}_{-0.06}$ & 0.99 (91.32/92) & $4.94\pm 0.26$ & $ 23.89 \pm 7.29 $ &  $ 2.41 \pm 1.47 $ \\	
		
		B & $0.12^{+0.02}_{-0.02}$ & $2.43^{+0.21}_{-0.20}$ & $0.79^{+0.19}_{-0.12}$ & 0.74 (74.68/101) & $5.70\pm 0.28$ & $ 16.04 \pm 4.35 $ & $ 1.59 \pm 0.86 $\\
		
		C & $0.12^{+0.02}_{-0.02}$ & $2.54^{+0.26}_{-0.25}$ & $0.90^{+0.42}_{-0.20}$ & 1.16 (91.67/79) & $5.65\pm 0.25$ & $ 14.16 \pm 6.14 $ & $ 1.41 \pm 1.22 $\\
		
		D & $0.09^{+0.02}_{-0.03}$ & $2.27^{+0.36}_{-0.32}$ & $0.69^{+0.32}_{-0.13}$ & 1.13 (64.25/57) & $4.77\pm 0.29$ & $ 18.77 \pm 8.73 $ & $ 1.90 \pm 1.77 $\\
		
		E & $0.15^{+0.02}_{-0.02}$ & $2.73^{+0.38}_{-0.30}$ & $0.87^{+0.74}_{-0.21}$ & 1.17 (91.60/78) & $5.31\pm 0.21$ & $ 13.23 \pm 8.78 $ & $ 1.19 \pm 1.58 $\\
		
		F & $0.10^{+0.02}_{-0.02}$ & $2.03^{+0.19}_{-0.18}$ & $0.60^{+0.07}_{-0.06}$ & 1.32 (188.99/143) & $5.13\pm 0.16$ & $ 23.32 \pm 5.16 $ & $ 2.56 \pm 1.13 $\\
		
		G & $0.13^{+0.01}_{-0.01}$ & $2.57^{+0.18}_{-0.17}$ & $0.81^{+0.17}_{-0.11}$ & 1.00 (178.80/178) & $5.68\pm 0.15$ & $ 14.79 \pm 3.43 $ & $ 1.39 \pm 0.64 $\\
		
		H & $0.10^{+0.02}_{-0.02}$ & $2.27^{+0.28}_{-0.27}$ & $0.65^{+0.17}_{-0.10}$ & 1.06 (111.01/105) & $5.11\pm 0.32$ & $ 19.38 \pm 6.35 $ & $ 1.91 \pm 1.25 $\\
		
		I & $0.13^{+0.02}_{-0.02}$ & $2.22^{+0.29}_{-0.28}$ & $0.60^{+0.14}_{-0.08}$ & 1.00 (77.99/78) & $5.45\pm 0.39$ & $ 20.8 \pm 6.66 $ & $ 2.03 \pm 1.3 $\\
		
		J & $0.08^{+0.02}_{-0.02}$ & $2.08^{+0.29}_{-0.28}$ & $0.56^{+0.12}_{-0.08}$ & 0.85 (78.81/93) & $5.00\pm 0.35$ & $ 23.44 \pm 7.93 $ & $ 2.41 \pm 1.63 $\\
		
		K & $0.14^{+0.02}_{-0.02}$ & $2.36^{+0.25}_{-0.24}$ & $0.63^{+0.12}_{-0.08}$ & 1.15 (112.25/98) & $5.44\pm 0.30$ & $ 18.79 \pm 4.93 $ & $ 1.74 \pm 0.91 $\\

	\hline
	\end{tabular}%
	\egroup
	\begin{tablenotes}\scriptsize
\item[*] In units of $10^{-13}$~erg~cm$^{-2}$~s$^{-1}$.
\end{tablenotes}
	\end{threeparttable}
\end{table}

For optically thick thermal Comptonization, the relation between the power-law spectral index ($\alpha = \Gamma-1$), electron temperature ($kT_e$), and optical depth ($\tau$) can be written as \citep{zdziarski1996},
\begin{equation}
\alpha = \left[\frac{9}{4}+\frac{1}{(kT_e/m_ec^2)\tau(1+\tau/3)}\right]^{1/2}-\frac{3}{2}.
\label{eq:tau}
\end{equation}
Equation~\ref{eq:tau} can be inverted to obtain the optical depth of the Comptonizing medium, given the photon index and plasma temperature. Using the best-fit spectral parameters from Table~\ref{nthcomp_fit}, the optical depth of the electrons lies in the range $13.2-23.9$ over the span of observations, which justifies our assumption of optically thick plasma. The Compton $y$ parameter in the optically thick limit is given by $4kT_e\tau^2/m_ec^2$, which quantifies the degree of modification of the seed photon spectrum by Compton scattering. The $y$ parameter for our fits vary in the range $1.2-2.6$, which imply significant alteration of the incident photon energy by the electron plasma. The last two columns of Table~\ref{nthcomp_fit} list the optical depth and $y$ parameter of the spectral fits.

\section{Discussion}
\label{sec:discuss}

\par We have presented here the results of our extensive analysis of eleven \textit{XMM-Newton} observations of magnetar source CXOU~J$010043.1-721134$ spanning over 2000 to 2016. We have shown the period evolution of this magnetar and calculated its revised spin-period and spin-down rate from long-term observations. We have found a double-peaked pulse profile, similar to that found by TE08, and we argue that a single-component model is better physically motivated to explain its emission. We have fitted its PN and MOS spectra with various models and present the thermally Comptonized continuum model as the best description of the physical processes at work giving rise to the emission from this magnetar.

\par As of December~2016, \mgr$\,$ is found to have a spin period of 8.0275(1)~s, consistent with the few-second periods seen for all magnetars. In fact, low-period X-ray pulsations are one of the distinguishing characteristics of magnetars, which result from rapid braking due to magnetic dipole radiation from their millisecond birth periods \citep{duncan1992} according to the magnetar model. \mgr$\,$ is seen to steadily spin-down from October~2000 to December~2016, with $\dot{P} = (1.76\pm 0.02) \times 10^{-11}$~s~s$^{-1}$. This value is similar to those calculated by MG05 and TE08 ($1.88\times 10^{-11}$ and $1.9\times 10^{-11}$~s~s$^{-1}$ respectively) for this source. The inferred characteristic age ($P/2\dot{P}$) of \mgr$\,$ is then $\sim 7200$~yr, consistent with the 6800~yr age derived by MG05. Magnetars, typically aged $10^3-10^4$ years, hence are extremely young objects. 

\par Unlike \mgr , not all magnetars exhibit steady spin-down over large timescales. For instance, the AXP~CXOU~J$171405.7-381031$ located in the supernova remnant CTB~37B shows erratic spin-down behaviour. \citet{gotthelf2019} monitored this magnetar over a 8-year span and found that its spin-down rate doubled in $< 1$~year before settling to a stable value of $5\times 10^{-11}$~s~s$^{-1}$. On the other hand, some AXPs exhibit significant spin-up and spin-down glitches following periods of radiative bursts or long-term flux enhancements (e.g. 1E~$1048.1-5937$ \citealt{gavriil2004, archibald2020}, 1E~$2259+586$ \citealt{woods2004}). Of the eight known persistent AXPs, \mgr$\,$ is one of the four (the other three being 1RXS~J$170849.0-400910$, CXOU~J$171405.7-381031$ and PSR~J$1622-4950$) from which X-ray bursts have not yet been observed. Although current observations seem to imply \mgr$\,$ to be a persistent magnetar showing no signs of bursting activities, possibilities of unobserved outbursts and/or glitches cannot be ruled out. For example, \citet{dib2014} estimate the approximate timescale of significant radiative change (e.g. flux outburst) to be at least several years for AXPs.

\par In the oblique rotator model for pulsars \citep{pacini1967}, the rotational kinetic energy of the neutron star provides the energy emitted by it due to magnetic dipole radiation. Surface dipolar magnetic field, $B$, is hence related to the dynamic properties of the star, viz. period ($P$) and period derivative ($\dot{P}$) as \citep{michel1991},
\begin{equation}
B=3.2\times 10^{19} \sqrt{P\dot{P}}\,\,\mathrm{G}.
\end{equation}
For \mgr, this gives a dipolar field strength of $3.8\times 10^{14}$~G. The value is in agreement with that of majority of magnetars \citep{olausen2014} which have magnetic field strengths $\sim$ a few times $10^{14}$~G and in some cases $> 10^{15}$~G (e.g. SGR~$1806-20$ \citealt{palmer2005}). The spin-down luminosity of this source ($\dot{E}\equiv 4\pi^2 I\dot{P}/P^3$, taking stellar moment of inertia $I\simeq 10^{45}$~g~cm$^2$) comes out to be $\sim 1.3\times 10^{33}$~erg~s$^{-1}$, which is two orders of magnitude lower than its observed X-ray luminosity ($\sim 10^{35}$~erg~s$^{-1}$). This excess energy is believed to be derived from the twisting and decay of its internal magnetic field according to the magnetar model.  

\par The double-peaked pulsed profile seen for \mgr$\,$ is not uncommon in magnetars. Indeed, such profiles have been obtained earlier for \mgr$\,$ by TE08 as well as for other magnetars such as CXOU J164710.2$-$455216 \citep{an2019}, XTE J1810$-$197 \citep{camilo2016}, Swift~J1822.3$-$1606 \citep{scholz2012} and so on. Unlike \mgr$\,$ which does not exhibit any significant differences in pulse profile in the two energy bands, some AXPs such as 1RXS~$170849-400910$ have pulse profile morphologies that are strongly energy-dependent \citep{denhartog2008}. The pulse fraction $(32\pm 3)\%$ calculated by TE08 in the $0.2-6$~keV energy range falls in the lower end of the range of pulse fraction $(29-63)\%$ computed by us for the broader peak of \mgr. In contrast, some AXPs have pulse fractions which are energy dependent. For example, the pulse fraction of 1E~$1048.1-5937$ increases with energy upto 8~keV and then decreases at higher energies \citep{yang2016}.

\par Magnetar spectra have been traditionally fitted with the phenomenological blackbody plus power-law model or sometimes with double blackbody model of two different temperatures. As mentioned in Section~\ref{subsec:spec} for a double blackbody model, the cooler blackbody dominates the emission in the soft band whereas in the hard band, the hotter blackbody dominates. We have also shown in Section~\ref{subsec:res_time} that the pulse profile in both energy bands are very similar. Two hotspots of different temperatures on the surface of the magnetar (each contributing majorly in separate energy bands) giving rise to the same pulse profile in both bands is unlikely. Hence, even though two-component models can successfully fit its spectra, we argue that to explain the observed pulse profiles, the modulation in both bands are more likely to have a common origin, resulting from and tied to a single physical process. 

\par Among the several single-component models used by us to fit the spectra, the thermal Comptonization model is favored by goodness-of-fit as well as physical considerations. The Comptonizing medium for thermal seed photons from the surface (with assumed $kT_{\mathrm{BB}}=0.2$~keV) is a cool ($kT_{e,\mathrm{avg}}\sim 0.7$~keV), optically thick ($\tau_{\mathrm{avg}}\sim 18.8$) electron plasma. Interestingly, the plasma temperature does not seem to decrease significantly with time, as would be expected if one were to assume a continuously radiating body with no heating sources. From the average electron temperature and luminosity of the source over the observation span, one can derive an approximate plasma cooling time ($t_{\mathrm{cool}}\sim N\langle kT_e\rangle/\langle L\rangle$, where $N \sim n_eV_{\mathrm{scat}}$ and $n_e \sim \langle\tau\rangle/l_{\mathrm{scat}}\sigma_T$) after making a few simplifying assumptions. Adopting a cylindrical scattering volume $V_{\mathrm{scat}}$ of height $l_{\mathrm{scat}}=1$~km and radius of an equivalent circle having the same area as that of the fraction of the total neutron star surface which is emitting (assumed to be 0.8), the cooling timescale is of the order of a few ms. Varying the assumed parameters within a reasonable range of values does not modify the order of magnitude of the cooling time. This implies that there exists an internal source within the system which continually provides heat to maintain its temperature over such long timescales.  

\par To explain the observed double-peaked pulse profiles, we propose a simple two-hotspot model for the neutron star, with the hotspots connected or coupled in such a way as to ensure the sameness of their temperatures. The similarity of pulse fractions of the two peaks also puts constraints on the size of the two hotspots -- for example, one hotspot cannot be significantly bigger than the other, and neither can occupy too significant a portion of the magnetar surface.  A detailed model for the two hotspots regarding their size, position on the surface as well as their orientation geometry with respect to the rotation axis and the observer, which attempts to reproduce the observed pulse profiles, is currently under investigation and will be presented elsewhere.

\par For completeness, we also note here that some magnetars display a significant spectral up-turn above $\sim$10~keV \citep{vogel2014} and the bulk of their flux comes from above 10~keV. To check for signatures of up-turn at higher energies for \mgr, one needs to examine its spectral data from \textit{NuSTAR}-like observatories, although such observations have not been made for this source yet.

\section{Summary}
\label{sec:summary}

Based on our in-depth analysis of \textit{XMM-Newton} observations of \mgr, we summarize our results as follows,
\begin{enumerate}
\item \mgr$\,$ is a persistent magnetar with a spin-period of 8.0275(1)~s (as of December~2016), and it exhibits steady spin-down with a spin-down rate of $(1.76\pm 0.02) \times 10^{-11}$~s~s$^{-1}$. This translates to a dipolar magnetic field of $3.8\times 10^{14}$~G and characteristic age of $7200$~years. These results are in agreement with previous studies (see MG05 and TE08).
\item It has a double-peaked pulse profile, and the profile shows no significant energy-dependence, consistent with the results earlier obtained for this magnetar. The pulse fractions in the soft and hard energy band lie in the range $(19-47)\%$ and $(18-63)\%$ respectively. The pulse profile as well as pulse fraction do not show notable temporal variations.
\item The thermal Comptonization model is able to adequately explain the observed spectra and pulse-profile. A cool, optically thick electron plasma acts as the Comptonizing medium for thermal seed photons from the magnetar surface. The source shows subtle spectral variations over the 16-years span of observations.
\end{enumerate}

\section*{Acknowledgements}
The authors are thankful to the reviewer whose comments and suggestions have helped to improve the quality of this manuscript. This research is based on observations obtained with \textit{XMM-Newton}, an ESA science mission with instruments and contributions directly funded by ESA Member States and NASA. The authors thank GH, SAG; DD, PDMSA and Director, URSC for encouragement and continuous support to carry out this research.

\section*{Data Availability} 
Data underlying this work is available at High Energy Astrophysics Science Archive Research Centre (HEASARC) facility, located at NASA-Goddard Space Flight Centre.


\bibliographystyle{mnras}
\bibliography{References} 

\bsp	
\label{lastpage}
\end{document}